\documentclass[runningheads]{llncs}
\usepackage{babel}
 
\usepackage{graphicx}
\usepackage{algorithm}
\usepackage{algpseudocode}

\usepackage{amsfonts}
\usepackage{mathtools}
\usepackage{amssymb}
\usepackage{bbm}
 
\usepackage{color} 
\usepackage{hyperref}
 
\usepackage[normalem]{ulem}

 \usepackage{cancel}
\usepackage{geometry}
\usepackage{subcaption}

\usepackage{nicematrix}
\usepackage{booktabs}

\usepackage[colorinlistoftodos, obeyDraft]{todonotes} 

\makeatletter

\usepackage{tikz}
\usetikzlibrary{patterns}
\usetikzlibrary{arrows.meta,positioning,fit,backgrounds,decorations.pathreplacing}

\makeatother

\newcommand{\Order}{\mathsf{Order}}
\newcommand{\CP}{\mathsf{CP}}

\newcommand{\Vote}{\mathsf{Vote}}
\newcommand{\Ptr}{\mathsf{Predecessor}}

\newcommand{\TxCert}{\mathsf{TC}}

\usepackage{algorithm}
\usepackage{algpseudocode}

\title{How Not to Be Censored: Inclusion Safety in Client-Led Consensus\thanks{The authors are listed in alphabetical order. We thank Roger Wattenhofer, Jovan Komatovic, and Tobias Klenze for helpful comments and suggestions.}}
\titlerunning{How Not to Be Censored}
\author{Yackolley Amoussou-Guenou\inst{1} \and Andrea Canidio\inst{2}}
\authorrunning{Y. Amoussou-Guenou and A. Canidio}
\institute{Universit\'e Paris-Panth\'eon-Assas, CRED, Paris, France \\ ~\\\and Category Labs, New York, USA. \email{acanidio@category.xyz} \\(corresponding author)}

\begin{document}

\maketitle

\begin{abstract}
Blockchain consensus protocols implicitly follow a client-server architecture: validators are the only active participants, while clients (i.e., users) merely submit transactions and rely on validators to include them. This abstraction obscures a fundamental question: what should a client do to guarantee the timely inclusion of its own transaction? We study this question by introducing a leaderless consensus protocol in which clients are active. We define \emph{inclusion safety}: the guarantee that a client's transaction enters some of the blocks produced for an epoch, and that every block omitting it belongs to a fork that is eventually abandoned. Because every transaction carries a reference to a block of the previous epoch, clients help select the surviving fork. This yields two paths to inclusion: reaching sufficiently many honest replicas within an epoch, or relaying a block containing the transaction across epochs until the corresponding fork becomes canonical. In a generalized model spanning synchrony and asynchrony, the client's and the replicas' connectivity are substitutes for inclusion safety: it is achieved either if the client is well connected, or if the client is poorly connected but honest replicas communicate reliably, provided the client reaches at least one of them. However, the substitution is imperfect: when inclusion safety rests on the replicas' connectivity, it requires a longer epoch compared to when the client is well connected.
\end{abstract}

\section{Introduction}

Canonical consensus protocols adopt an implicit client-server architecture. Validators (i.e., the replicas that run the protocol) understand its rules, may or may not follow them, and are rewarded or penalized depending on their behavior. Clients (i.e., users) are instead passive and naive: they submit transactions without knowing how the protocol works and rely on validators to include them in a block. Because clients are assumed to do little, most models of consensus leave them out entirely: they begin from one or more proposers that already hold a set of transactions, and rarely ask where those transactions came from.

This assumption matters most for censorship resistance, that is, whether and how a blockchain can guarantee timely inclusion of valid transactions. The client-server architecture creates a thorny principal-agent problem: clients must rely on validators to include their transactions. Yet in most models the problem cannot even be posed: clients are absent. The assumption of passive clients is also unrealistic. Many clients are controlled by professional organizations (e.g., financial institutions, arbitrageurs, searchers, solvers) whose profits depend on understanding how the protocol works. And when ``naive'' users submit transactions, they do so through other professional organizations (e.g., wallets, RPC endpoints, builders) whose business depends on understanding the protocol's rules and reacting to them.

The main contribution of this paper is to study censorship resistance \textit{from the client's point of view}. We introduce a new property, \textit{inclusion safety}: the guarantee that a client's transaction is included in some of the blocks produced for a given epoch, and that every block omitting it belongs to a fork that is eventually abandoned. Unlike the classical properties of consensus, inclusion safety is specific to a transaction: it depends on the protocol, but also on the behavior of the client that submits it. We therefore study inclusion safety in a novel consensus protocol that is client-led, in the sense that clients play an active role in it, and we derive conditions --- on the protocol's parameters and on the client's ability to communicate --- under which it holds.

More precisely, we present a leaderless consensus protocol in which every replica collects transactions from clients and proposes them for inclusion in the next block, which is constructed by merging a quorum of those proposals. As usual, fewer than one-third of the replicas are Byzantine and behave arbitrarily; the rest are honest and follow the protocol. Byzantine clients are unrestricted in number. Each epoch may produce several blocks: quorums keep forming as client proposals keep arriving. Crucially, the blocks produced for the same epoch are totally ordered. The protocol is client-led in a specific sense: every transaction a client submits carries a reference to a block of the previous epoch. When replicas merge proposals into a block, they set its predecessor to the highest referenced block they have observed. Through this reference, clients help select which fork of the chain survives.

We call a client \emph{compliant} if it broadcasts its transaction to all replicas, and the transaction is valid against the block it references and against any higher block. 
We show that a compliant client has two paths to inclusion safety, one within an epoch and one across epochs. On the first path, the transaction reaches sufficiently many honest replicas in time, and every block produced for that epoch includes it. On the second path, the transaction reaches only a few replicas and enters only some blocks of the epoch. The client then carries a block containing it forward, epoch after epoch, until the reference reaches sufficiently many honest replicas. From that point on, every surviving block descends from a block containing the transaction, while blocks that omit it belong to forks that are eventually abandoned.

We study the protocol's properties --- and inclusion safety in particular --- in two network environments. The first is a generalized network model that spans the synchronous and the asynchronous case. The client's communication technology is represented by the set of replicas it reaches reliably, that is, with latency below a known bound. The replicas' communication technology is represented by the subset of honest replicas that reach each other reliably. All other messages can be delayed arbitrarily, though not indefinitely. We show that liveness always holds, while safety and inclusion safety depend on both technologies. Furthermore, the client's and the replicas' communication technologies are substitutes: a well-connected client achieves inclusion safety independently of inter-replica communication, while a poorly connected client still achieves it when honest replicas communicate reliably, provided it reliably reaches at least one of them. The substitution is however imperfect: when the client is poorly connected, information travels through an intermediate replica, and inclusion safety requires a longer epoch --- twice as long, in our model.

In this environment, however, the second path plays no role: for every parameter value, either the first path succeeds and the transaction enters all blocks of the epoch, or the attacker prevents inclusion altogether. The intuition is that an attacker can relay messages arbitrarily fast. If too few honest replicas observe the client's transaction in time to include it in their proposals, the attacker can assemble a quorum that excludes it, use the resulting block to open the next epoch, and delay every other quorum until the submission period of the subsequent epoch has closed: no block of the epoch then includes the transaction. If instead enough honest replicas include the transaction in their proposals, every possible quorum contains it, and every block of the epoch includes it.

To study the second path, we therefore turn to a second environment, in which latency is random rather than adversarial. There, with positive probability, a transaction enters some blocks of an epoch but not others, and the second path becomes active: the client relays a block containing its transaction forward, until every active fork contains a block that includes it. We quantify by simulation how the epoch length trades off inclusion within an epoch against recovery across epochs.

The rest of the paper proceeds as follows. The remainder of this section discusses the related literature. Section~\ref{sec:model} introduces the model, and Section~\ref{sec:problem} defines inclusion safety. Section~\ref{sec:protocol} describes the protocol. Section~\ref{sec:properties} contains our main results: liveness, safety, and inclusion safety in the generalized network model, and the analysis of the second path to inclusion under stochastic latency. Section~\ref{sec:conclusion} concludes. All mathematical derivations missing from the text are in Appendix~\ref{sec:math-appendix}.

\subsection*{Related literature}

Our protocol belongs to a recent literature studying protocols in which several replicas can submit transactions for inclusion at the same time, i.e., \emph{multiple concurrent proposer} (MCP) protocols \cite{Avarikioti2023fnfbft,garimidi2025multipleconcurrentproposers,elsheimy2026censorship,babel2026cadence} or \emph{leaderless} protocols  \cite{antoniadis2021leaderless,keidar2021dag,danezis2022narwhal,xiang2026prefixconsensuscensorshipresistant}. As is common in these protocols, ours can produce several blocks for the same epoch, and the blocks of an epoch are comparable: each can be read as a prefix of the higher ones.

This literature engages directly with censorship resistance: with several proposers, a transaction has several independent routes into the chain. But because clients are absent from these models, the definitions refer to replicas: censorship resistance holds if a transaction held by an honest replica is guaranteed inclusion on chain --- immediately, in the strongest formulations \cite{babel2026cadence,garimidi2025multipleconcurrentproposers}, or within a bounded delay \cite{elsheimy2026censorship}. The guarantee reaches the client only through unstated assumptions --- for example, that the client's transaction has reached at least one honest replica in time.\footnote{A partial exception is \cite{abraham2025latency}, whose \emph{input holder} is the analogue of our client. There, however, the input holder communicates synchronously with every other party, so inclusion is guaranteed as long as it is honest. The same is true in our model when the network is fully synchronous; our goal is to characterize inclusion safety when it is not.} Our work is concerned with precisely this missing step: we place the client in the model and ask which client behavior secures inclusion.

A parallel line of work strengthens censorship resistance while keeping a single leader, by constraining the leader's behavior. In inclusion-list designs, a committee of replicas compiles lists of pending transactions that the next block must contain: FOCIL \cite{thiery2024focil} enforces the lists through the fork-choice rule and requires only one honest committee member, at a latency cost --- lists are compiled during the preceding slot, so enforced inclusion arrives one block later. AUCIL \cite{wadhwa2025aucil} designs the list-construction game so that rational parties build informative lists. In these designs, too, the guarantee is stated at the level of committee members, which the client's transaction must reach.

A related literature studies the economics of censorship resistance. \cite{fox2023censorship} define it as the cost an attacker must bear to exclude a transaction, given the fee it pays; \cite{alpos2025overview} apply this lens to deployed protocols, ranking them by the bribes required to delay inclusion; \cite{stouka2025feemechanism} ask which transaction fee mechanisms make censorship expensive. This literature shares our starting point --- what secures the inclusion of a given transaction --- but we follow the classical approach of distributed computing and assume Byzantine replicas, which censor at any price. Our model is therefore more conservative; inclusion safety is nonetheless achievable, not by paying more, but by communicating better. Finally, \cite{sridhar2025adversarymajority} show that the modeling of clients --- whether always online, whether communicating --- resolves discrepancies among resilience bounds, but do not analyze transaction inclusion; in our work, the client acts to secure the inclusion of its own transaction.

\section{Model}\label{sec:model}

The system has two types of processes: \emph{replicas} and \emph{clients}. Replicas are what the blockchain literature calls validators or miners; we denote their set by $R$, with $|R|=n$. Clients are the users of the system and have transactions to submit; their number is finite.

Time is divided into periods called \emph{epochs}. For each process, an epoch is one run of the protocol's algorithm. When epoch $e$ ends, epoch $e+1$ begins. Epochs need not be aligned across processes. Each replica runs the same algorithm once per epoch: the algorithm takes as input one or more blocks from the previous epoch that the replica has observed, and returns one or more blocks for the current epoch, each pointing to a single block from the previous epoch. 

\paragraph{Failure model.}
Some processes may deviate from the prescribed protocol. A process that deviates is called \emph{Byzantine}; a process that follows the protocol is called \emph{correct} or \emph{honest}. At most $f$ replicas are Byzantine, with $n=3f+1$. We place no restriction on the number of Byzantine clients.

\paragraph{Communication model.}
We consider a generalized network model, in which synchrony is a property of individual links rather than of the whole network: the client communicates reliably with a set of replicas, and a subset of the honest replicas communicate reliably with each other. Communicating reliably means delivering messages within a known bound $\Delta$, while communicating non-reliably means that messages can be delayed arbitrarily, though not indefinitely. The model includes the classical synchronous and asynchronous models as special cases, and is inspired by granular synchrony \cite{giridharan2024granular}, which also treats synchrony as a property of individual links; unlike that model, ours distinguishes the client's links from the replicas' links.  Section~\ref{sec:properties} describes the model in detail and derives the protocol's properties as functions of the two communication technologies.

Each message carries an unforgeable digital signature, and no process can alter a message already sent. When a process delivers a message, it knows which process created it. We assume a broadcast primitive with the following guarantees: when an honest process broadcasts a message, every honest process eventually delivers it; and if one honest process delivers a message, every honest process eventually delivers it. Here \emph{delivery} is the local event by which a process accepts a message and hands it to the protocol; it is the point at which the protocol's rules react to the message, and it may occur later than the message's physical receipt. A Byzantine process may modify the delivery delay of its own messages.

\paragraph{Blocks and predecessors.}
A \emph{block} is a set of transactions together with a pointer to a single block from the previous epoch, called its \emph{predecessor}. We write $\mathsf{pred}(b)$ for the predecessor of block $b$. Execution starts from a distinguished \emph{genesis block}, the unique block of epoch $0$; every block of epoch $1$ has the genesis block as its predecessor. Because each block points to a single predecessor, the blocks produced across all epochs form a tree rooted at the genesis block: a directed acyclic graph in which every block has exactly one parent. The next two definitions describe this structure.

\begin{definition}[Predecessor chain]
The \emph{predecessor chain} of a block $b$ produced in epoch $e$ is the sequence of blocks obtained by following predecessor pointers back to the genesis block:
\[
b_0,\,b_1,\,\dots,\,b_{e-1},\,b_e,
\qquad b_e=b,\quad b_{s-1}=\mathsf{pred}(b_s)\ \text{for } 1\le s\le e .
\]
We say that $b$ \emph{extends} a block $b'$ if $b'$ lies on the predecessor chain of $b$.
\end{definition}

A predecessor chain is one candidate history of the system: one block per epoch, each extending the one before. Several such histories may coexist.

\begin{definition}[Active forks]
Let $\mathcal{B}_e$ be the set of blocks produced in epoch $e$. The set of \emph{active forks} at epoch $e$ is the set of predecessor chains of the blocks in $\mathcal{B}_e$:
\[
\mathcal{F}_e=\bigl\{\,\text{predecessor chain of } b \;:\; b\in\mathcal{B}_e\,\bigr\}.
\]
\end{definition}

In general, two blocks of $\mathcal{B}_e$ may have predecessor chains that agree up to some earlier epoch and then branch apart, so $\mathcal{F}_e$ may hold several histories \cite{antonella_blockchainADT}. However, when an epoch produces a single block, it has a single active fork, which we call the canonical chain.\footnote{Our use of \emph{canonical chain} differs slightly from the usual one, in which the canonical chain is the chain selected by some rule --- the longest chain in Bitcoin, the heaviest-attested chain in Ethereum. Here no selection is needed: the chain is canonical because it is unique.}

\section{Problem definition}\label{sec:problem}

We study the problem of transaction inclusion, in which clients have an active role. To do so, we need to introduce a notion of transaction validity. 

Each block induces a state: the state of its predecessor after the block's transactions are applied. A transaction is \emph{valid against} a block $b$ if it can be applied (or executed) to the state that $b$ induces. Note that validity is distinct from successful execution. For example, in the context of Ethereum, a transaction transferring ERC20 tokens that the sender does not own is valid --- it can be applied and included in a block --- but it reverts at execution. By contrast, a transaction with an incorrect nonce or an insufficient native balance cannot be applied at all: it is invalid. Validity is also state-dependent, so the same transaction may be valid against one block of an epoch and invalid against another: if a client spends funds in some block of epoch $e-1$, a transaction spending those same funds is invalid against that block but may be valid against another block of the same epoch that omits the earlier spend.


\begin{definition}[Inclusion safety]\label{def:inclusion-safety}
Let $x$ be a transaction submitted for epoch $e$, and let $\tilde{\mathcal{B}}_{e-1}(x)\subseteq \mathcal{B}_{e-1}$ be the set of blocks in $\mathcal{B}_{e-1}$ against which $x$ is valid. The protocol \emph{achieves inclusion safety for $x$} if $\tilde{\mathcal{B}}_{e-1}(x)$ is non-empty and the following hold.
\begin{itemize}
    \item \textbf{Immediate inclusion.} At least one block of epoch $e$ with predecessor in $\tilde{\mathcal{B}}_{e-1}(x)$ contains $x$.
    \item \textbf{Invalid not included.} No block of epoch $e$ with predecessor in $\mathcal{B}_{e-1} \setminus \tilde{\mathcal{B}}_{e-1}(x)$ contains $x$.
    \item \textbf{Eventual agreement.} There is an epoch $e^\star\ge e$ such that, for every epoch $e'\ge e^\star$, every active fork in $\mathcal{F}_{e'}$ contains a block that includes $x$.
\end{itemize}
\end{definition}

Two points about this definition. First, it separates inclusion from agreement. Inclusion is immediate: the transaction appears in some block of the epoch in which it is submitted. Agreement may take longer: histories that omit the transaction may be pruned epoch by epoch, until every active fork carries it --- not delayed inclusion, but immediate inclusion with possibly delayed agreement.

Second, inclusion safety is specific to a transaction submitted \textit{for a given epoch}. As we will see, it can hold only if the client knows when the submission window for that epoch is open. Conversely, a malicious actor can break inclusion safety by manipulating this window.

Finally, throughout, we also assume that blocks are never full: block space is not scarce, and a block can accommodate every admissible transaction. With scarcity, a transaction could be excluded from a block even when all the conditions for inclusion safety below are met --- for example, because other transactions pay higher fees

\section{Protocol description}\label{sec:protocol}

The execution of the protocol proceeds in epochs $e = 1,2,\ldots$. During each epoch, clients submit proposals, replicas exchange information about the proposals they have observed, and replicas vote on views of the epoch to achieve some agreement. A quorum of votes forms a \emph{vote certificate}. A vote certificate together with a known and deterministic ordering rule defines a \emph{certified block}.

An epoch may produce multiple vote certificates, as different replicas may certify different blocks; the protocol guarantees that all vote certificates of an epoch are totally ordered (Lemma~\ref{lem:vc-chain}).

Clients drive progress from one epoch to the next. Upon receiving a vote certificate from epoch $e-1$, a client enters epoch $e$ and broadcasts a proposal containing a transaction together with a vote certificate. 

\begin{figure}[t]
\centering
\resizebox{\textwidth}{!}{
\begin{tikzpicture}[
    x=1cm, y=1cm,
    round/.style={draw=blue!55!black, fill=blue!12, minimum height=0.75cm, font=\small, inner sep=2pt},
    bnd/.style={dotted, thick, black!70},
    mlabel/.style={font=\scriptsize, align=center}
]
\draw[->, thick] (-0.4,1.7) -- (13.4,1.7) node[right, font=\small] {time};

\draw[thick] (0.2,1.9) -- (0.2,1.5);
\node[mlabel, anchor=south] at (0.2,2.0) {epoch $e$\\begins};
\draw[thick] (12.6,1.9) -- (12.6,1.5);
\node[mlabel, anchor=south] at (12.6,2.0) {epoch $e$\\ends};

\node[round, minimum width=3.0cm, anchor=west] (r1) at (0.2,1.0)  {Client proposals};
\node[round, minimum width=3.2cm, anchor=west] (r2) at (3.2,1.0)  {Replica proposals};
\node[round, minimum width=6.1cm, anchor=west] (r3) at (6.4,1.0)  {Witness round};

\node[round, minimum width=4.4cm, anchor=west] (r4) at (8.2,0.1) {Vote round};

\draw[bnd] (3.2,1.55) -- (3.2,0.55);
\node[mlabel, anchor=north, text=blue] at (3.2,0.25) {close\\boundary ($A$)};

\draw[bnd] (6.4,1.55) -- (6.4,0.55);
\node[mlabel, anchor=north, text=blue] at (6.4,0.25) {first close proposal \\ received};

\draw[bnd] (8.2,1.55) -- (8.2,-0.35);
\node[mlabel, anchor=north, text=blue] at (8.2,-0.65) {voting begins\\($2A$)};

\node[mlabel, anchor=north, text width=4.0cm] at (11.5,-0.65)
  {rounds overlap: each newly arrived close proposal yields a larger witness and a new vote};
\end{tikzpicture}
}
\caption{\small The rounds of one epoch, as seen by a single replica. The epoch begins when the replica receives its first valid client proposal and ends when a vote certificate triggers the next epoch. The replica collects client proposals until its close boundary, at time $A$ after the epoch opens, then broadcasts its close proposal. It collects close proposals for a further interval $A$ and starts voting at time $2A$: it builds a support witness from the close proposals observed so far and votes on it. The witness and vote rounds overlap: close proposals keep arriving after the first vote, and each one lets the replica build a larger witness and vote again.}
\label{fig:epoch-timeline}
\end{figure}
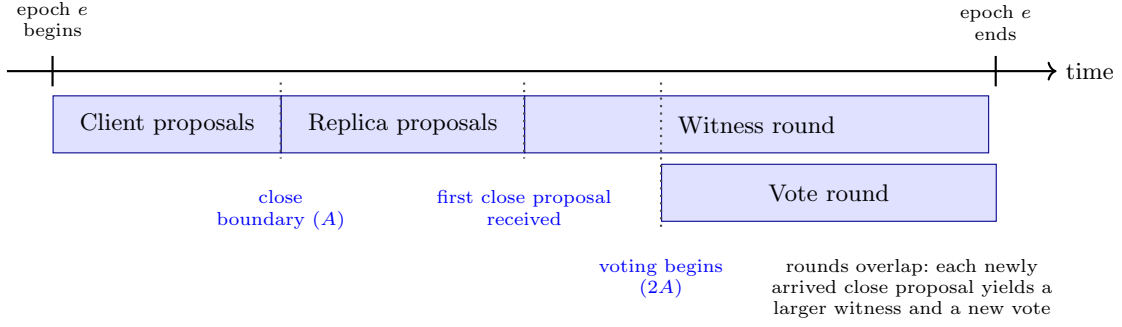

In more detail, each epoch consists of three rounds: client proposal, witness construction, and voting (illustrated in Figure~\ref{fig:epoch-timeline}).

\paragraph{Client Proposals}

In this round, clients broadcast their proposal for epoch $e$. 

A client proposal for epoch $e$ is a tuple
$P=(c,e,tx,\mathsf{meta},VC(W_{e-1}))$, 
where $c$ is the client identifier, $tx$ is the (possibly empty) transaction payload,
$\mathsf{meta}$ contains transaction metadata, and
$VC(W_{e-1})$ is a vote certificate from epoch $e-1$.

When receiving a proposal, a replica verifies summarily whether it is valid, i.e., whether the attached vote certificate $VC(W_{e-1})$ is correctly signed and the transaction is valid relative to the state induced by the certified block corresponding to $VC(W_{e-1})$; in that case, it re-broadcasts the proposal to the other replicas.
A replica opens epoch $e$ as soon as it receives its first valid proposal for that epoch. 

Each replica maintains a local timer. Let $A$ denote the \emph{close-boundary} parameter.
Once a replica opens epoch $e$, it accepts valid client proposals arriving before the timer expires. At expiration, the replica broadcasts a signed \emph{close proposal}, $CP_i(e)$, representing all valid client proposals it has delivered. If replica $i$ never receives a valid proposal for epoch $e$ but observes at least $f+1$ close proposals from other replicas, it broadcasts an empty close proposal.

\paragraph{Witness Construction}

Replicas then aggregate the close proposals they observe into \emph{support witnesses}.


Whenever a replica observes two distinct close proposals from the same
replica for the same epoch, it flags that replica as equivocating.
Similarly, if two distinct client proposals from the same client are observed
for the same epoch, that client is flagged as equivocating.

Formally, a support witness built by replica $i$ during epoch $e$ is: $W_{i,e}=(\mathcal P,\Ptr,F,NF)$,
where $F$ is the set of flagged replicas; $NF$ is the set of non-flagged replicas from which a close proposal has been received; $\mathcal P$ is the set of valid client proposals contained in the close proposals of replicas in $NF$; $\Ptr$ is the highest certified predecessor referenced by any observed close proposal.
      

A witness grows as the replica observes additional close proposals, in one of three elementary ways. Call 
$
W_{i,e} = \bigl(\mathcal{P},\,\Ptr,\,F,\, NF\bigr),
$
the initial witness and 
$
W'_{i,e} = \bigl(\mathcal{P}',\,\Ptr',\,F',\, NF'\bigr),
$
the witness constructed after observing an additional close proposal. If the additional close proposal is from a new replica, we have $F' = F$, $NF \subseteq NF'$, and $\mathcal{P} \subseteq \mathcal{P}'$. If the additional close proposal is from a non-flagged replica for which an earlier close proposal was observed, this replica is flagged and we have $NF' \cup F' = NF \cup F$, $F \subseteq F'$, and $\mathcal{P}' \subseteq \mathcal{P}$. Finally, if it comes from a replica already in
$F$, the sets $\mathcal{P}$, $F$, and $NF$ are unchanged. In all three cases
$\Ptr \preceq \Ptr'$, since the set of observed close proposals only grows. 

The fact that witnesses grow as additional close proposals are observed provides a natural way to rank them: we say that $W \preceq W'$ if $W'$ can be obtained from $W$ by observing finitely many additional close proposals. We also write $W \prec W'$ if $W \preceq W'$ and $W \neq W'$. 

To create a block, the replica must remove some proposals from its witness. Given a witness, replica $i$ removes all proposals that are invalid relative
to $\Ptr$ together with all proposals from equivocating clients.
Let $S(W_{i,e})$ denote the resulting set of admissible transactions. The
candidate block associated with the witness is
$B(W_{i,e}) = \bigl( \Order(S(W_{i,e})), \Ptr(W_{i,e}) \bigr),$
where $\Order(\cdot)$ is a deterministic ordering function. With a slight abuse of notation, we also write $B(W_{i,e})\preceq B(W'_{i,e})$ if $W_{i,e}\preceq W'_{i,e}$.
\paragraph{Voting}
Each replica $i$ maintains a local lock $L_i$ initially set to the empty set. 

Replica $i$ may broadcast a signed message $\Vote_i(e,W_{i,e})$, i.e., vote for its locally built witness $W_{i,e}$  whenever $|NF \cup F| \ge 2f+1$
and either
$L_i=\emptyset
$
or
$
L_i \prec W_{i,e}.
$
After voting, replica $i$ updates its lock, setting $L_i$ to $W_{i,e}$.

Because witnesses may grow over time, an honest replica may cast multiple votes during an epoch, but only for a strictly increasing sequence of witnesses.

Replica $i$ begins voting only after waiting an additional interval $A$ after its close boundary, that is, at time $2A$ after opening the epoch.

\paragraph{Vote Certificates and Certified Blocks}

When receiving votes from at least $2f+1$ different replicas for the same witness $W_e$ during epoch $e$, the replica aggregates them into a \emph{vote certificate}  $VC(W_e)$. 
If a replica has a vote certificate for a witness $W_e$, it considers the associated block as a \emph{certified block}.


Multiple vote certificates can be produced in a single epoch. A crucial property is that all vote certificates of a given epoch form a chain (Lemma~\ref{lem:vc-chain}). 
A replica stops voting for epoch $e$ when the close boundary of an epoch $e' > e$ starts.

\section{Properties}\label{sec:properties}

An epoch may produce multiple vote certificates (and thus multiple certified blocks). The next lemma shows that these certificates are totally ordered.
\begin{lemma}[Vote certificates form a chain]
\label{lem:vc-chain}
If $VC(W_e)$ and $VC(W'_e)$ are two different vote certificates in epoch $e$, then either $W_e \prec W'_e$ or $W'_e \prec W_e$.
\end{lemma}

We now analyze the properties of the protocol under different network conditions. In particular, we study liveness, safety, and inclusion safety. Many of these properties depend on the behavior of clients, which we formalize as follows.

\begin{definition}[Compliant client]
A client is \emph{compliant} if, in a given epoch $e$:
\begin{itemize}
    \item  It broadcasts a single valid proposal to all replicas (delivery depends on network conditions);
    \item (\emph{Monotone validity}) the transaction in its proposal is valid relative to any certified block from epoch $e-1$ that ranks higher than the one included in its proposal.
\end{itemize}
\end{definition}

\noindent
Intuitively, a compliant client maximizes the likelihood of inclusion by disseminating its transaction --- and, with it, a recent certified block --- as widely as possible; monotone validity ensures the transaction stays valid if the protocol later selects a higher-ranked predecessor. We deliberately avoid the term \emph{honest}: correctness does not depend on client behavior, and compliance is about qualifying for the inclusion guarantees below.

Monotone validity rules out strategies in which a client exploits forks by submitting conflicting transactions across epochs: a second transaction reusing a nonce, or spending the same funds, may be valid relative to its chosen block yet invalid relative to a higher-ranked certified block that includes the earlier transaction. Monotone validity excludes exactly this behavior. 

Monotone validity is operationally relevant at the filtering step defining $S(W_e)$: when constructing a candidate block from a witness $W_e$, replicas discard any proposal whose transaction is invalid relative to the predecessor pointer $\Ptr(W_e)$. By definition, a valid proposal contains a transaction that is valid relative to the certified block from epoch $e-1$ included in the proposal. If such a proposal is submitted by a compliant client, by monotone validity the included transaction remains valid relative to any higher-ranked certified block, and in particular relative to $\Ptr(W_e)$ --- and, since $\Ptr$ is monotone, relative to $\Ptr(W'_e)$ for every $W'_e \succeq W_e$.

A useful consequence of the definition of a compliant client is the monotonicity of inclusion of its transaction within a fixed epoch.
\begin{lemma}[Same-epoch monotonicity]
\label{lem:same-epoch-monotonicity}
Let $B(W_e)$ and $B(W'_e)$ be two certified blocks from epoch $e$ such that $B(W_e)\preceq B(W'_e)$. Suppose $W_e$ is constructed using a close proposal from an honest replica containing a proposal from a compliant client. Then both $B(W_e)$ and $B(W'_e)$ contain the compliant client's transaction.
\end{lemma}

\noindent
A more subtle property is that this monotonicity extends across epochs.

\begin{lemma}[Cross-epoch monotonicity]
\label{lem:cross-epoch-monotonicity}
Let $s\ge 1$. Suppose that $B(W_{e+s})$ extends $B(W_e)$ and that
\[
B(W'_{e+s}) \succeq B(W_{e+s}).
\]
Then $B(W'_{e+s})$ extends some certified block $B(W'_e)$ such that
\[
B(W'_e) \succeq B(W_e).
\]
\end{lemma}

\begin{figure}[t]
\centering
\resizebox{.9\textwidth}{!}{
\begin{tikzpicture}[
    >={Latex[length=2.4mm]},
    block/.style={draw, rounded corners, align=center, minimum width=3cm, minimum height=0.95cm, font=\small},
    green/.style={block, fill=green!14, draw=green!50!black, thick},
    grayb/.style={block, fill=black!6, draw=black!40},
    ptr/.style={->, dashed, thick, draw=black!55},
    mono/.style={->, thick, draw=green!50!black},
    elabel/.style={font=\small\bfseries}
]
\node[elabel] at (0,3.6) {epoch $e-1$};
\node[elabel] at (4.5,3.6) {epoch $e$};
\node[elabel] at (9,3.6) {epoch $e+1$};

\node[green] (we) at (4.5,2.4) {$B(W'_e)$\\{\color{blue}$W_e\preceq W'_e$}};

\node[green] (a) at (0,0.8) {$B(W_{e-1})$\\contains $tx$};
\node[green] (b) at (4.5,0.8) {$B(W_e)$\\contains $tx$};
\node[green] (c) at (9,0.8) {$B(W_{e+1})$\\contains $tx$};

\node[grayb] (ga) at (0,-1.2) {$\widehat B_{e-1}$\\omits $tx$};
\node[grayb] (gb) at (4.5,-1.2) {$\widehat B_{e}$\\omits $tx$};
\node[grayb] (gc) at (9,-1.2) {$\widehat B_{e+1}$\\omits $tx$};

\draw[ptr] (b.west) -- (a.east);
\draw[ptr] (c.west) -- node[above, font=\scriptsize] {$\Ptr$} (b.east);
\draw[ptr] (we.west) -- (a.east);
\draw[ptr] (gb.west) -- (ga.east);
\draw[ptr] (gc.west) -- (gb.east);

\draw[mono] (b.north) -- node[right, font=\scriptsize] {preserves $tx$} (we.south);

\node[font=\scriptsize, align=center, text width=13cm] at (4.5,-2.9)
 {Dashed arrows are predecessor pointers. Green nodes are certified blocks containing the compliant client's transaction $tx$; gray nodes omit it. Once $tx$ appears, every higher or extending certified block preserves it.};
\end{tikzpicture}
}
\caption{\small Certified blocks form a DAG across epochs. Predecessor pointers define edges. Once a compliant client's transaction appears on a certified path, monotonicity implies that all higher certified extensions of that path preserve the transaction.}
\label{fig:certified-block-dag}
\end{figure}
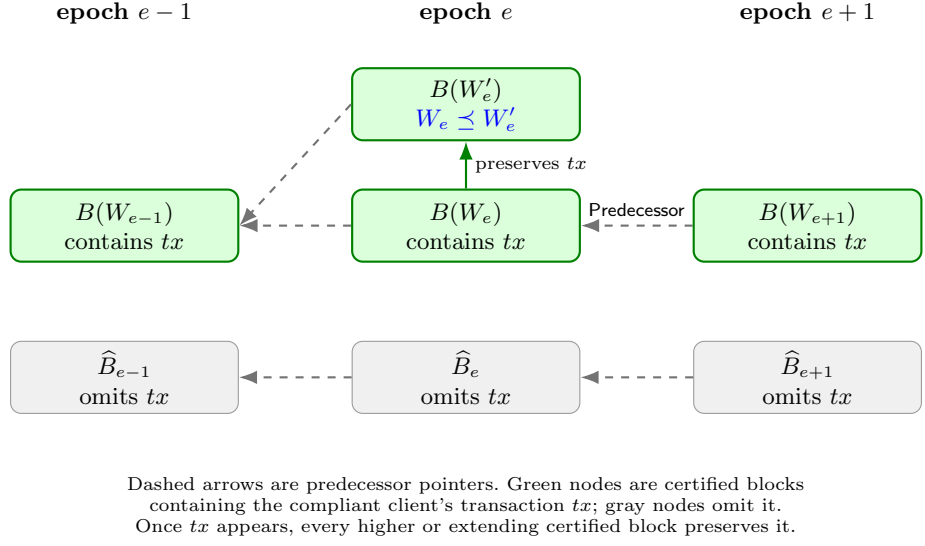

\noindent
The main implication of the above lemmas is that once a compliant client's transaction appears in a certified block via an honest replica, any higher certified block --- within the same epoch or across epochs --- extends a chain that preserves that inclusion (see Figure~\ref{fig:certified-block-dag}).

We now present a sufficient condition for inclusion of a transaction in all active forks at a given epoch that is independent of the network assumptions. For each honest replica $i$, let $\mathcal{S}_{e,i}$ denote the set of proposals contained in the close proposals of \emph{honest} replicas (including $i$ itself) that $i$ observes before casting its first vote for epoch $e$. Unlike the full set $\mathcal{P}$, which can shrink when a Byzantine sender is flagged, this part persists in every witness that $i$ subsequently votes for. Also, the set $\mathcal{S}_{e,i}$ is an analytical device, because honest replicas cannot identify other honest replicas.

A proposal enters $\mathcal{S}_{e,i}$ if $i$ observes it ``early enough'': before its voting period starts (no earlier than $2A$ after opening, and only after observing close proposals from $2f+1$ distinct replicas), or within the close proposals of the first $2f+1$ distinct replicas. The next lemma shows that if sufficiently many honest replicas observe a compliant client's proposal early enough, then the proposal is included in every certified block for that epoch.
\begin{lemma}[Inclusion safety]
\label{prop:inclusion-safety}
Let $P$ be a proposal from a compliant client for epoch $e$, and let
\[
H_P = \{i \text{ honest} : P \in \mathcal{S}_{e,i}\}.
\]
If $|H_P|\ge f+1$, then:
\begin{itemize}
    \item \textbf{Same-epoch inclusion.} The transaction $\mathsf{tx}(P)$ is included in every certified block $B(W_e)$.

    \item \textbf{Cross-epoch persistence.} Let $B(W_{e-1})$ be the certified block included in $P$. Then every certified block $B(W_e)$ satisfies
    \[
    \Ptr(W_e)\succeq B(W_{e-1}).
    \]
\end{itemize}
\end{lemma}

\begin{figure}[t]
\centering
\resizebox{.8\textwidth}{!}{
\begin{tikzpicture}[
    x=1.1cm, y=0.8cm,
    >={Latex[length=2.2mm]},
    dotL/.style={circle, fill=green!60!black, inner sep=0pt, minimum size=4pt},
    dotM/.style={circle, fill=blue!70!black, inner sep=0pt, minimum size=4pt},
    dotH/.style={circle, fill=red!75!black, inner sep=0pt, minimum size=4pt},
    lineL/.style={green!60!black, very thick},
    lineM/.style={blue!70!black, very thick},
    lineH/.style={red!75!black, very thick}
]
\fill[green!10] (0.3,0) rectangle (8.2,5);

\draw[dashed, thick, blue!40!black] (0,5) -- (8.3,5);
\node[anchor=south west, font=\scriptsize, blue!45!black] at (0.15,5.12) {rank $2f{+}1$};
\draw[dashed, thick, black!45] (5,0) -- (5,8.0);
\node[anchor=south, font=\scriptsize, black!60] at (5,8.05) {$2f{+}1$-th replica};

\draw[->, thick, black!70] (0,0) -- (8.9,0);
\draw[->, thick, black!70] (0,0) -- (0,8.6);
\node[anchor=north, font=\scriptsize] at (4.2,-0.95) {replicas, ordered by how early they observe the proposal};
\node[rotate=90, anchor=south, font=\scriptsize] at (-0.7,4) {rank at which the proposal is first observed};

\foreach \i in {1,...,7} {\node[anchor=north, font=\scriptsize, black!70] at (\i,-0.12) {\i};}

\draw[lineL] (1,1) -- (2,1) -- (3,2) -- (4,2) -- (5,3) -- (6,3) -- (7,4);
\foreach \p in {(1,1),(2,1),(3,2),(4,2),(5,3),(6,3),(7,4)} {\node[dotL] at \p {};}
\node[anchor=west, font=\scriptsize, green!55!black] at (7.25,4) {$P_1$ (guaranteed)};

\draw[lineM] (1,1) -- (2,2) -- (3,3) -- (4,4) -- (5,5) -- (6,6) -- (7,6);
\foreach \p in {(1,1),(2,2),(3,3),(4,4),(5,5),(6,6),(7,6)} {\node[dotM] at \p {};}
\node[anchor=west, font=\scriptsize, blue!65!black] at (7.25,6) {$P_2$ (guaranteed)};

\draw[lineH] (1,3) -- (2,5) -- (3,6) -- (4,6) -- (5,7) -- (6,7) -- (7,7);
\foreach \p in {(1,3),(2,5),(3,6),(4,6),(5,7),(6,7),(7,7)} {\node[dotH] at \p {};}
\node[anchor=west, font=\scriptsize, red!70!black] at (7.25,7) {$P_3$ (not guaranteed)};

\draw[black!55, thick] (5,5) circle (3.4pt);
\end{tikzpicture}
}
\caption{\small Inclusion safety for three proposals ($f=2$ for illustration). For a given proposal, replicas are ordered along the horizontal axis by how early they observe it (earliest on the left), and a replica's height is the rank of the first close proposal of an honest replica that contains it, so a lower point means the proposal was seen earlier; each curve is sorted on its own. The shaded band is the first-vote cutoff: ranks at most $2f+1$. A proposal is included in every certified block of epoch $e$ when at least $2f+1$ replicas observe it within their first $2f+1$ close proposals, equivalently when its curve lies on or below the corner $(2f+1,2f+1)$; at least $f+1$ of those replicas are then honest. Observing the proposal before the voting period starts, at time $2A$, also qualifies regardless of rank; the picture shows the rank condition only. Proposals $P_1$ and $P_2$ satisfy this and are guaranteed, while $P_3$ rises above the cutoff before the $(2f+1)$-th replica and is not. }
\label{fig:inclusion-safety}
\end{figure}
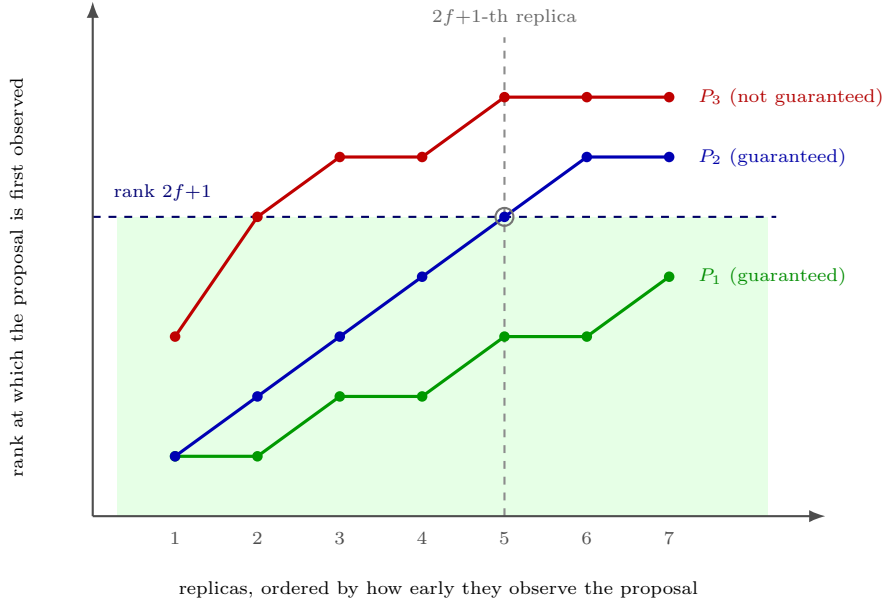

\noindent
The lemma highlights two distinct mechanisms for inclusion. The first operates within a single epoch: if a compliant client disseminates its proposal to sufficiently many honest replicas early enough, then its transaction is included in all certified blocks for that epoch. The second operates across epochs: by propagating a certified block whose chain contains the transaction, the client preserves it along all extensions, until it appears in all active forks. This second channel is \emph{not} delayed inclusion --- the transaction is already in some certified blocks of its epoch; what is delayed is agreement: blocks omitting it are progressively eliminated from the active forks (see Figure~\ref{fig:inclusion-safety}). 

Inclusion safety depends on sufficiently many honest replicas observing the proposal, through the close proposal of an honest replica, before they cast their first vote. Operationally, this occurs in two stages: first, the client disseminates its proposal to some replicas; second, those replicas further disseminate it by including it in their close proposals. The next subsection makes this two-step diffusion process explicit and derives conditions --- on the communication technology of clients and replicas --- under which Lemma~\ref{prop:inclusion-safety} applies and inclusion safety of a compliant client's transaction is guaranteed.

\subsection{Generalized network model}\label{sec:intermediate}

We study the properties of the protocol in a generalized network model that includes the traditional synchronous and asynchronous models as special cases. We do so for two reasons. First, the traditional models constrain only the communication between replicas, while in our setting the communication abilities of clients matter as well; any synchrony assumption must therefore be paired with an assumption on the clients' side. Second, an interesting result emerges: the clients' and the replicas' communication technologies are imperfect substitutes in achieving inclusion safety. Inclusion requires the client to both learn that the epoch has opened --- possibly through a hidden vote certificate --- and place its proposal in sufficiently many honest close proposals before the close boundaries. A directly connected client runs each leg in one hop; a poorly connected client runs each in two, through a relaying replica.

 Let $C$ be the set of replicas with which the client communicates reliably: messages between the client and a replica in $C$, in either direction, are delivered within a known bound $\Delta$. Let $R$ be a set of honest replicas that communicate reliably with each other, again within $\Delta$. Every other message is delivered after a delay chosen by the adversary, finite but unbounded. The adversary can also make any message arbitrarily fast. We write $H$ for the set of honest replicas, so that $R\subseteq H$, and treat $|C\cap H|$ and $|R|$ as measures of the client's and the replicas' connectivity, and $C\cap R$ as their overlap. Neither set is known to the client, which broadcasts its proposal to all replicas; connectivity determines which of its messages arrive in time, not where it sends them.

If $C=R=\emptyset$, the model is fully asynchronous; if $R=\emptyset$ but $C\neq\emptyset$, only the client's links to $C$ are reliable. Symmetrically, $C=R=H$ is the fully synchronous model, while $R=H$ with $C\cap H\subsetneq H$ leaves only the client's connectivity unreliable.

We first note that the protocol is live for every $C$ and $R$.
\begin{proposition}[Liveness]\label{prop:liveness}
Fix an epoch $e$ and suppose some honest replica receives a valid proposal for epoch $e$. Then, for every $C$ and $R$ --- including $C=R=\emptyset$ --- epoch $e$ produces at least one vote certificate.
\end{proposition}

In this model, a client's proposal can reach a replica in two ways: directly, if the replica is in $C$, or relayed by a replica in $C\cap R$ to the rest of $R$. Accordingly, define the client's \emph{reach} as
\[
N \;=\;
\begin{cases}
(C\cap H)\cup R, & \text{if } C\cap R\neq\emptyset,\\[2pt]
C\cap H, & \text{if } C\cap R=\emptyset .
\end{cases}
\]

\begin{proposition}[Inclusion Safety]
\label{prop:reach-dichotomy}
Fix an epoch $e$ and a compliant client that broadcasts its proposal upon first delivering a vote certificate of epoch $e-1$.
\begin{itemize}
    \item \textbf{Direct guarantee.} If $|C\cap H|\ge f+1$ and $A\ge 2\Delta$, the client's transaction is included in every certified block of epoch $e$.
    \item \textbf{Relayed guarantee.} If $C\cap R\neq\emptyset$, $|N|\ge f+1$ and $A\ge 4\Delta$, the same conclusion holds.
    \item \textbf{Impossibility.} If $|N|\le f$, then for every $A$ and every client strategy there exists a Byzantine strategy such that no certified block of epoch $e$ contains the client's transaction. In particular, inclusion safety fails, and relaying vote certificates across epochs cannot restore it.
\end{itemize}
\end{proposition}

\noindent
The threshold is the quorum-intersection count of Lemma~\ref{prop:inclusion-safety}: the guarantees work by placing the proposal in the own close proposal of every replica in $N$, and $|N|\ge f+1$ triggers the lemma. Conversely, when $|N|\le f$, the $f$ Byzantine replicas together with the honest replicas outside $N$ can certify a witness on their own. In this case, the attacker can also prevent the creation of any other vote certificate: it creates its preferred vote certificate arbitrarily fast, uses it to open the next epoch, and delays all other messages until enough honest replicas reach their close boundary for that epoch, which ends their voting phase for epoch $e$. This has an important implication: the second channel for inclusion is absent here. Either the transaction enters every certified block of the epoch, or the attacker ensures a single certified block that excludes it; partial inclusion never arises in this model.

The guarantees place the proposal in the close proposal of every replica in $N$, in two ways. When the client is well connected, $|C\cap H|\ge f+1$, the situation is essentially the synchronous one: one delay $\Delta$ to learn a (possibly hidden) vote certificate, another to be heard, so $A\ge 2\Delta$. When the client is poorly connected, the replicas make up the difference --- $|C\cap H|$ and $|R|$ must together account for $f+1$ replicas --- and the regime is slower: the chain has four links (the opening propagates within $R$, a replica in $C\cap R$ informs the client, the client broadcasts, that replica relays to the rest of $R$), so it requires $A\ge 4\Delta$. The substitution between the two connectivities is therefore imperfect: reaching more replicas directly halves the epoch length the guarantee demands.

We now turn to safety, that is, whether the certified blocks of epoch $e$ agree on their predecessor. The guarantee mirrors Proposition~\ref{prop:reach-dichotomy}: what matters is no longer the connectivity of a single client, but the joint connectivity of all compliant clients whose proposals reference the block in question.
\begin{proposition}[Safety]\label{prop:safety}
Fix an epoch $e$ and a certified block $M$ of epoch $e-1$. Consider the compliant clients whose proposal for epoch $e$ references the certified block $M$, each broadcasting it upon first delivering a vote certificate of epoch $e-1$. Let $\mathcal{C}_M$ be the union of their reliable sets, and define their joint reach
\[
N_M \;=\;
\begin{cases}
(\mathcal{C}_M\cap H)\cup R, & \text{if } \mathcal{C}_M\cap R\neq\emptyset,\\[2pt]
\mathcal{C}_M\cap H, & \text{if } \mathcal{C}_M\cap R=\emptyset .
\end{cases}
\]
\begin{itemize}
    \item \textbf{Direct guarantee.} If $|\mathcal{C}_M\cap H|\ge f+1$ and $A\ge 2\Delta$, every certified block $B(W_e)$ satisfies $\Ptr(W_e)\succeq M$.
    \item \textbf{Relayed guarantee.} If $\mathcal{C}_M\cap R\neq\emptyset$, $|N_M|\ge f+1$ and $A\ge 4\Delta$, the same conclusion holds.
    \item \textbf{Impossibility.} If $|N_M|\le f$, then for every $A$ there exists a Byzantine strategy such that some certified block satisfies $\Ptr(W_e)\prec M$.
\end{itemize}
\end{proposition}

Predecessor pointers may still differ across the certified blocks of epoch $e$: a close proposal revealed after the first votes may reference a block above $M$, which later certificates extend. All pointers are totally ordered, and under the guarantees any disagreement involves only blocks above $M$.

\subsection{Stochastic latency and the role of the second path to inclusion}\label{sec:alternating}

The analysis so far leaves the second path to inclusion --- relaying a vote certificate whose chain contains the transaction --- without a clear role: for every parameter value, either a compliant client's transaction is included in all blocks produced for an epoch, or there is no inclusion safety. The reason is that an attacker can arbitrarily speed up some messages, creating an early vote certificate, while slowing down all others so that no further certificate is produced: either the attack fails and every certificate includes the transaction, or it succeeds and none does. This section instead studies a model in which latency is random and not controlled by an adversary; there, a transaction may enter some certified blocks of an epoch but not all.

We modify the communication model as follows. Each message between honest processes is delivered after a delay drawn by nature, independently across messages and epochs, from a distribution $F$ with full support $[\underline{p},\bar{p}]$, where $0<\underline{p}\le\bar{p}<\infty$. Unlike in the previous sections, the adversary does not control delays: Byzantine processes choose the content and the timing of their \emph{sends}, but their messages, too, are delivered with delays drawn from $F$. This captures congestion and jitter rather than targeted attacks; $A$ must now be chosen against a random environment.

Similarly to the synchronous case, if $A\ge 2\bar{p}$  the first path to inclusion (i.e., inclusion in \textit{all} vote certificates for an epoch) is always active, and the second path is never needed. At the other extreme, a small $A$ makes inclusion unlikely, but not impossible. The client cannot reach any replica earlier than $2\underline{p}$ after the first honest opening: it needs at least $\underline{p}$ to learn a vote certificate and at least $\underline{p}$ more to be heard. Close boundaries are not aligned, however, and the dispersion works in the client's favor: a replica that opens the epoch late also closes late, and may therefore still receive the proposal in time. Since an opening can lag the first opening by at most $\bar{p}-\underline{p}$, inclusion in all certified blocks becomes impossible only when $A\le 3\underline{p}-\bar{p}$, a bound that is vacuous whenever $\bar{p}\ge 3\underline{p}$. In what follows we focus on the regime
\[
\max\{\underline{p},\,3\underline{p}-\bar{p}\} \;<\; A \;<\; 2\bar{p},
\]
where the first path succeeds with probability strictly less than one. 

\begin{lemma}[Second path to inclusion]
\label{lem:partial-inclusion}
Suppose $3\underline{p}-\bar{p}<A<2\bar{p}$ and $F$ is non-degenerate. Fix an epoch $e$ and a compliant client that broadcasts its proposal $P$ upon first delivering a vote certificate of epoch $e-1$.
\begin{itemize}
    \item[(i)] Suppose each Byzantine replica behaves like an honest replica, except that it omits $P$ from its close proposal and that it votes for every witness whose block omits $\mathsf{tx}(P)$. If $A<2\bar{p}-\underline{p}$, then, with strictly positive probability, epoch $e$ produces certified blocks $B(W^{*})\prec B(W^{**})$ with $\mathsf{tx}(P)\in B(W^{**})\setminus B(W^{*})$. If in addition a Byzantine client opens the epoch, withholding the opening proposal from one honest replica, the same conclusion holds for every $A<2\bar{p}$.
    \item[(ii)]Partial inclusion requires the adversary's participation. If the Byzantine replicas send no close proposals, then, with probability one, $\mathsf{tx}(P)$ is included either in every certified block of epoch $e$ or in none.
\end{itemize}
\end{lemma}

\noindent
Whenever some Byzantine replica supplies a close proposal omitting the transaction, it is therefore possible that a compliant client's transaction is included in only some blocks but not all. The client's proposal arrives before the close boundary at some honest replicas and after it at others. An early witness omitting it is certified first, and the epoch ends with a totally ordered set of certified blocks, split into a lower range that omits the transaction and an upper range that contains it.

It remains to show that relaying a vote certificate whose block contains the transaction achieves inclusion safety. Suppose a compliant client's transaction is contained in the certified block $B(W_{\hat e})$ of epoch $\hat e$. In each later epoch, upon first delivering a vote certificate of the previous epoch, the client relays, in an empty-payload proposal, the highest certificate it has delivered whose block extends $B(W_{\hat e})$. If it reaches sufficiently many honest replicas early enough --- the hypothesis of Lemma~\ref{prop:inclusion-safety}, which has positive probability in every epoch --- every certified block of that epoch extends a block carrying the transaction; by Lemmas~\ref{lem:same-epoch-monotonicity} and~\ref{lem:cross-epoch-monotonicity}, the transaction then belongs to every active fork, permanently: inclusion safety is achieved.

Whether the relay lands before the forks carrying the transaction are abandoned is a race governed by the epoch length $A$: raising $A$ postpones every close boundary and first vote, enlarging the set of delay realizations in which the relay succeeds. In Appendix~\ref{app:simulation} we simulate this race under a representative adversary that opens each epoch with a hidden vote certificate and censors the relay, with $f=2$ and delays drawn uniformly from $[\underline{p},\bar{p}]=[1,3]$. Two findings stand out. First, the probability that a relay epoch ends with the transaction in every certified block increases steeply with $A$: in our parametrization it is near one already at $A\approx 4$, well below the conservative bound $2\bar{p}=6$ at which the first path is always active. Second, relay epochs essentially resolve one way or the other, so a client falling back on the second path rarely waits more than one epoch. Below $2\bar{p}$, inclusion safety is therefore probabilistic, with a probability that rises quickly with the epoch length; certainty returns at $A\ge 2\bar{p}$.

\section{Conclusion}\label{sec:conclusion}

This paper studies censorship resistance from the client's point of view: we introduce a leaderless consensus protocol in which every transaction references a block of the previous epoch, and we define \emph{inclusion safety}, the guarantee that a client's transaction enters some of the blocks produced for an epoch, and that every block omitting it belongs to a fork that is eventually abandoned. We show that there are two paths to inclusion safety: reaching sufficiently many honest replicas within an epoch, or relaying a block containing the transaction across epochs until the corresponding fork becomes canonical. In a generalized network model, the clients' and the replicas' connectivity are imperfect substitutes for inclusion safety; under stochastic latency, the second path completes inclusion with a probability our simulations quantify.



\bibliographystyle{splncs04}
\bibliography{bib}

\appendix
\newpage
\section{Pseudocode}\label{sec:app_pseudo-code}

\begin{algorithm}[H]
\caption{Pseudo-code for replica $r$}
\label{alg:pseudo_code}
\scriptsize
\begin{algorithmic}[1]
\Statex Inputs: $e$ \Comment{Value of the current epoch} 
\Statex 
\State $  \textit{epoch} \gets e$
\State $VC_{\textit{epoch}}^{(r)} \gets \emptyset$
\Statex
\State  $CP \gets \emptyset$ \Comment{valid  proposals from client seen by the replica for epoch for the current epoch}
\State $CP\_seen( \textit{epoch}) \gets \emptyset$ \Comment{seen close proposals $(j,CP_j( \textit{epoch}))$ for each epoch $ \textit{epoch}$}
\State $Witness\_delivered_{\textit{epoch}} \gets \emptyset$


\Statex
    \Statex //Beginning of epoch  \textit{epoch}
    \State $CP \gets \emptyset$
    \State $CP\_seen( \textit{epoch}) \gets \emptyset$
    
\While{\textbf{true}}
    \State close-boundary timer set, but not started
    \Statex
    \Statex //waits for the first valid client proposal
    \Statex // (Client proposals round)

    \Statex
    \While{close-boundary timer not expired}
        \If {client proposal $P=(c, \textit{epoch},tx,\mathsf{data},VC(W_{ \textit{epoch}-1}))$ such that $\textsc{ValidProposal}(P,W_{ \textit{epoch}-1})$ deliverd}

    \State Start close-boundary timer
    
    \While{close-boundary timer not expired}
        \State upon receiving $P'=(c', \textit{epoch},tx',\mathsf{data}',VC(W'_{ \textit{epoch}-1}))$:
        \If{$\textsc{ValidProposal}(P', W'_{ \textit{epoch}-1})$}
            \State $CP \gets CP \cup \{(P',  \textit{epoch})\}$
        \EndIf
    \EndWhile
    
        \EndIf
        \If {close-boundary timer not started and $|\{j : (j, CP_j) \in CP\_seen(e')| \ge f+1 : e'\ge  \textit{epoch}$ delivered}
            \State $ \textit{epoch} \gets e'$ 
            \State $CP \gets \emptyset$
            \State Set close-boundary timer to expire
        \EndIf
    \EndWhile    
%
   
    \Statex

    \State \label{line:bc-CP}broadcast $\CP$ to all replicas 
    
    \Statex
    \Statex // (Witness and Vote round)
    \Statex
    
    \State $W_{i,\textit{epoch}}= \emptyset$
    \State $L_{\textit{epoch}}= \emptyset$
    
    \While{$VC(W_{\textit{epoch}})= \emptyset$} 

	
        \State $all\_replicas\_seen(\textit{epoch}) = \{r' | \exists CP' \text{ created by r' } : CP' \in CP\_seen(\textit{epoch})\}$
        \State  $equivocating\_replicas(\textit{epoch}) = \{r' | flag(r', \textit{epoch}) = faulty\_on\_CP\}$
        \State $non\_equivocating\_replicas(\textit{epoch}) = all\_replicas\_seen(\textit{epoch}) \setminus equivocating\_replicas(\textit{epoch})$

        \State $valid\_proposals(e) = \{P | (P, \textit{epoch}) \in non\_equivocating\_replicas(\textit{epoch}\}$

        \State $\Ptr = Pred\_highest\_certified\_block(CP\_seen(\textit{epoch}))$ 
        \State $W_{i,\textit{epoch}} = (valid\_proposals(\textit{epoch}),\,\Ptr,\,equivocating\_replicas(\textit{epoch}),\, non\_equivocating\_replicas(\textit{epoch}))$

	\Statex
    \If{ $|all\_replicas\_seen(\textit{epoch})| \ge 2f+1$}

        \If{  $L_{\textit{epoch}} \prec W_{i,\textit{epoch}}$}
        		\State broadcast $\Vote(\textit{epoch}, W_{i,\textit{epoch}})$
        		\State $L_{\textit{epoch}} \gets W_{i,\textit{epoch}}$
        \EndIf
        
        \Statex
        \If{  $\exists W \in Witness\_delivered_{\textit{epoch}} : |\Vote(\textit{epoch},W) \ge 2f+1$}

        		\State $VC(W_{\textit{epoch}})= W$
        \EndIf


    \EndIf
    
    \EndWhile
    
    \Statex
    \State $VC_{e}^{(r)} \gets \textsc{HighestRelevantVC}(e)$
    \State \textit{Creation of the block}
    \State \textbf{epoch $e$ ends} 
    \State $e \gets e+1$
    \Statex
\EndWhile
\end{algorithmic}
\end{algorithm}

\begin{algorithm}[H]
\caption{Communication handler}
\label{alg:communication}
\scriptsize
\begin{algorithmic}[1]

\While{\textbf{true}} \Comment{Loop handling the reception of messages}
  \If{reception of $(\emph{msg}, epoch, j)$} \Comment{epoch is the epoch associated to the message, $j$ is the initiator of the message}
	\If{\textsf{type}(\emph{msg}) = client proposal}
		\If{ $\exists (msg', j) \in setMessages(epoch)$}
			\State $flag(j, epoch) \gets faulty\_client$
		\EndIf
		\If{$epoch$ > current\_epoch} \Comment{the current\_epoch in which the process is}
			\State \textbf{go to} epoch $epoch$
		\EndIf
	\Else \If{\textsf{type}(\emph{msg}) = CP}
			\If{$\exists (msg', j) \in CP\_seen(epoch)$}
				\State $flag(j, epoch) \gets faulty\_on\_CP$
			\EndIf
			\State $CP\_seen(epoch) \gets CP\_seen(epoch) \cup msg$

			\If{$CP\_seen(epoch) \ge f+1$ and $epoch > current\_epoch$} \Comment{current\_epoch in the variable representing the epoch in which the process is}
				\State \textbf{go to} epoch $epoch$
			\EndIf
		
		\Else
			\If{\textsf{type}(\emph{msg}) = Witness}
				\State $Witness\_delivered_{\textit{epoch}} \gets Witness\_delivered_{\textit{epoch}} \cup msg$
			\EndIf
		 
			\State setMessages $\gets$ setMessages $\cup$ \emph{msg}
		\EndIf
	\EndIf
  	  \EndIf
\EndWhile

\end{algorithmic}
\end{algorithm}

\section{Mathematical Appendix}\label{sec:math-appendix}

\subsection{Proof of Lemma~\ref{lem:vc-chain}}
Each vote certificate contains $2f+1$ votes from distinct replicas. Since $n = 3f+1$, any two such sets intersect in at least $f+1$ replicas, and since at most $f$ replicas are Byzantine, at least one of these is honest.

An honest replica votes only for witnesses it has itself built, and by the lock rule each vote after the first is for a witness strictly above its previous one. Since $\preceq$ is transitive, any two witnesses it votes for are comparable. The two certificates share at least one honest voter, so either $W_e \prec W'_e$ or $W'_e \prec W_e$.

\subsection{Proof of Lemma~\ref{lem:same-epoch-monotonicity}}

Let $P$ be the client's proposal and $h \in NF(W_e)$ an honest replica whose close proposal contains $P$. By Lemma~\ref{lem:vc-chain}, $W_e \preceq W'_e$ implies that either $W_e = W'_e$  or $W'_e$ is obtained from $W_e$ by elementary steps.  Along elementary steps, replicas never leave $NF \cup F$, and $h$ never enters $F$. Hence $h \in NF(W'_e)$, its close proposal belongs to $W'_e$, and $P \in \mathcal{P}(W'_e)$.

Since $P \in \mathcal{P}(W'_e)$, we have $\Ptr(W'_e) \succeq \mathsf{block}(P)$, and monotone validity implies that $\mathsf{tx}(P)$ is valid relative to $\Ptr(W'_e)$. Also, the client is compliant and client signatures are unforgeable, so $\mathcal{P}(W'_e)$ contains no second distinct proposal from it. Hence $P$ is not removed when the block is constructed, so $\mathsf{tx}(P) \in S(W'_e)$ and therefore $\mathsf{tx}(P) \in B(W'_e)$. The claim for $B(W_e)$ is the degenerate case $W'_e = W_e$.

\subsection{Proof of Lemma~\ref{lem:cross-epoch-monotonicity}}
We prove the claim by induction on $s$.

\paragraph{Base case ($s=1$).}
Suppose that $B(W_{e+1})$ extends $B(W_e)$, i.e., $\Ptr(W_{e+1}) = B(W_e)$. Assume
\[
B(W'_{e+1}) \succeq B(W_{e+1}).
\]
Since both blocks are certified in epoch $e+1$ and $B(W'_{e+1}) \succeq B(W_{e+1})$, we have $W_{e+1} \preceq W'_{e+1}$, and monotonicity of $\Ptr$ gives
\[
\Ptr(W_{e+1}) \preceq \Ptr(W'_{e+1}).
\]
It follows that $\Ptr(W'_{e+1})$ is a certified block from epoch $e$ such that
\[
\Ptr(W'_{e+1}) \succeq B(W_e).
\]
By definition, $\Ptr(W'_{e+1})$ lies in the predecessor chain of $B(W'_{e+1})$, which proves the claim.

\paragraph{Induction step.}
Assume the statement holds for some $s\ge 1$. We prove it for $s+1$.

Suppose that $B(W_{e+s+1})$ extends $B(W_e)$ and that
\[
B(W'_{e+s+1}) \succeq B(W_{e+s+1}).
\]
Since both blocks are certified in epoch $e+s+1$, we have $W_{e+s+1} \preceq W'_{e+s+1}$, and monotonicity of $\Ptr$ gives
\[
\Ptr(W_{e+s+1}) \preceq \Ptr(W'_{e+s+1}).
\]

Let
\[
\Ptr(W_{e+s+1}) = B(W_{e+s}), \qquad \Ptr(W'_{e+s+1}) = B(W'_{e+s}).
\]
Then $B(W_{e+s})$ extends $B(W_e)$, and
\[
B(W'_{e+s}) \succeq B(W_{e+s}).
\]
By the induction hypothesis, $B(W'_{e+s})$ extends some $B(W'_e)$ such that
\[
B(W'_e) \succeq B(W_e).
\]

Finally, since $\Ptr(W'_{e+s+1}) = B(W'_{e+s})$ lies in the predecessor chain of $B(W'_{e+s+1})$, it follows that $B(W'_e)$ also lies in that chain. Hence $B(W'_{e+s+1})$ extends $B(W'_e)$.

\subsection{Proof of Lemma~\ref{prop:inclusion-safety}}

Consider any certified block $B(W_e)$. Its vote certificate contains $2f+1$ votes from distinct replicas; since $|H_P| \ge f+1$ and there are $3f+1$ replicas in total, some replica $r \in H_P$ voted for $W_e$. Honest replicas vote only for witnesses they build, so $r$ built $W_e$.

Since $P \in \mathcal{S}_{e,r}$, replica $r$ observed, before casting its first vote, a close proposal containing $P$ from some honest replica $h$ (possibly $r$ itself). Replica $h$ signs a single close proposal per epoch and signatures are unforgeable, so $r$ never flags $h$: in every witness that $r$ votes for, $h \in NF$ and $P \in \mathcal{P}$. In particular, $P \in \mathcal{P}(W_e)$.

The client is compliant and client signatures are unforgeable, so $\mathcal{P}(W_e)$ contains no second distinct proposal from it. Moreover, $P \in \mathcal{P}(W_e)$ implies $\Ptr(W_e) \succeq \mathsf{block}(P) = B(W_{e-1})$, and monotone validity implies that $\mathsf{tx}(P)$ is valid relative to $\Ptr(W_e)$. Hence $P$ is not removed when the block is constructed, so $\mathsf{tx}(P) \in S(W_e)$ and therefore $\mathsf{tx}(P) \in B(W_e)$. This proves same-epoch inclusion and the inequality $\Ptr(W_e) \succeq B(W_{e-1})$, which is cross-epoch persistence. The final claim follows because such a transaction belongs to every certified block of its epoch, and every predecessor chain contains one certified block per epoch.

\subsection{Proof of Proposition~\ref{prop:liveness}}
Let $i$ be an honest replica that receives a valid proposal for epoch $e$. Upon receipt, $i$ re-broadcasts it. Since all delays are finite, every honest replica eventually receives a valid proposal (or at least $f+1$ close proposals) and therefore eventually opens the epoch and broadcasts a close proposal. After waiting $2A$ from its opening, each honest replica may vote.

We show that honest observations converge. Each honest replica observes exactly one close proposal from each honest replica, and flags a Byzantine replica upon observing its second distinct close proposal, so the set of close proposals ever counted by honest replicas is finite. Honest votes are broadcast and carry their witnesses, and witnesses contain the underlying close proposals (or retrievable commitments to them); hence every close proposal observed by one honest replica --- including the pairs that evidence equivocation --- is eventually observed by all. Every honest replica's observations therefore converge to a common final view, with the same non-flagged senders $NF^{*}$, flags $F^{*}$, proposal set $\mathcal{P}^{*}$, and pointer $\Ptr^{*}$. The corresponding witness $W^{*} = (\mathcal{P}^{*},\Ptr^{*},F^{*},NF^{*})$ contains close proposals from the $2f+1$ honest replicas, and is therefore votable.

Finally, each honest replica's view evolves by elementary steps, so its lock --- a witness it previously built --- always lies below its current view under $\preceq$. Hence each honest replica eventually votes for $W^{*}$, and these $2f+1$ votes form a vote certificate.

\subsection{Proof of Proposition~\ref{prop:reach-dichotomy}}
\textbf{Direct guarantee.} A vote certificate requires close proposals from $2f+1$ distinct replicas, and a replica that never opens epoch $e$ sends none. Since $|C|\ge|C\cap H|\ge f+1$, the replicas outside $C$ number at most $3f+1-(f+1)=2f$, so every certificate of epoch $e$ contains the close proposal of a replica in $C$; in particular, if no replica in $C$ opens the epoch, no certificate forms and the claim holds vacuously. Otherwise let $\sigma$ be the first time at which a replica in $C$ opens epoch $e$, so that every replica in $C$ has its close boundary at $\sigma+A$ or later. That replica re-broadcasts the valid proposal that opened it, and its link to the client is bounded, so the client first delivers a vote certificate of epoch $e-1$ at some $t\le\sigma+\Delta$ and broadcasts its proposal then. Every replica in $C$ receives it by $t+\Delta\le\sigma+2\Delta\le\sigma+A$, hence before its own close boundary, and includes it in its own close proposal. Each replica in $C\cap H$ observes its own close proposal, so the client's proposal belongs to $\mathcal{S}_{e,i}$; since $|C\cap H|\ge f+1$, Lemma~\ref{prop:inclusion-safety} gives inclusion in every certified block.

\textbf{Relayed guarantee.} Since $N\subseteq C\cup R$, we have $|C\cup R|\ge f+1$, so by the counting above every certificate of epoch $e$ contains the close proposal of a replica in $C\cup R$; otherwise no certificate forms. Let $\sigma$ be the first time at which a replica in $C\cup R$ opens the epoch, so every replica in $C\cup R$ closes at $\sigma+A$ or later, and fix $x\in C\cap R$. If the first such replica lies in $C$, the client is informed within $\Delta$ as above; if it lies in $R$, its re-broadcast opens every replica in $R$ by $\sigma+\Delta$, and $x$ then informs the client by $\sigma+2\Delta$. Either way the client broadcasts its proposal by $\sigma+2\Delta$, so every replica in $C$ receives it by $\sigma+3\Delta$, and $x$ in particular. Replica $x$ re-broadcasts it, and since $x\in R$, every replica in $R$ receives it by $\sigma+4\Delta$. As $A\ge4\Delta$, all these arrivals precede the respective close boundaries, so every replica in $(C\cap H)\cup R=N$ carries the proposal in its own close proposal, and $|N|\ge f+1$ gives the conclusion by Lemma~\ref{prop:inclusion-safety}.

\textbf{Impossibility.} Suppose $|N|\le f$ and consider the coalition consisting of the $f$ Byzantine replicas and the honest replicas outside $N$, of which there are at least $2f+1-|N|$; the coalition has at least $2f+1$ members. Every link between the client and an honest coalition member is adversarial, as is every link between a coalition member and a replica in $N$: an honest coalition member is neither in $C$ nor, if $C\cap R\neq\emptyset$, in $R$. A Byzantine client opens the epoch for the coalition only, delivering a valid proposal with zero delay; all messages toward $N$, and all messages to and from the compliant client, are delayed until the coalition's voting has ended. The coalition's honest members exchange close proposals promptly, each observes a common set of $2f+1$ coalition close proposals containing its own, waits until time $2A$ after opening, builds that witness and votes for it, and the Byzantine votes complete a certificate whose block omits the client's transaction. The Byzantine client then opens epoch $e+1$ for the coalition, ending its voting phase for epoch $e$, and only afterwards are the replicas in $N$ opened. Whatever the client then does --- submitting a transaction or relaying a vote certificate --- a witness containing its proposal can gather at most the $|N|\le f$ votes of $N$, so no such witness is certified. All delays are finite.

\subsection{Proof of Proposition~\ref{prop:safety}}

\textbf{Direct guarantee.} Fix a client in the family and let $C_c$ be its reliable set. As in the proof of Proposition~\ref{prop:reach-dichotomy}, let $\sigma_c$ be the first time a replica in $C_c$ opens epoch $e$; the client first delivers a vote certificate of epoch $e-1$ at some $t\le\sigma_c+\Delta$ and broadcasts its proposal then. Every replica in $C_c$ receives it by $t+\Delta\le\sigma_c+2\Delta\le\sigma_c+A$, hence before its close boundary (a proposal arriving before the opening itself opens the epoch). By the definition of the family, this proposal references $M$. Taking the union over the family, every replica in $\mathcal{C}_M\cap H$ includes in its own close proposal a proposal referencing $M$. Now consider any certified block $B(W_e)$ and an honest replica $r$ among its voters: $r$ built $W_e$, and by the voting rule $|NF\cup F|\ge 2f+1$, so at most $f$ replicas lie outside $NF\cup F$. Since $|\mathcal{C}_M\cap H|\ge f+1$, some carrier $h\in\mathcal{C}_M\cap H$ lies in $NF\cup F$; $h$ is honest and signs a single close proposal per epoch, so $r$ never flags it and $h\in NF$. Its close proposal belongs to $W_e$, and since $\Ptr$ is computed over all observed close proposals, $\Ptr(W_e)\succeq M$.

\textbf{Relayed guarantee.} Let $c$ be a client in the family with $C_c\cap R\neq\emptyset$ and let $x\in C_c\cap R$. As in the proof of Proposition~\ref{prop:reach-dichotomy}, the epoch opens somewhere in $C_c\cup R$ at time $\sigma$; within $\Delta$ the opening propagates through $R$, within $2\Delta$ replica $x$ informs the client, within $3\Delta$ the client's proposal reaches $x$, and within $4\Delta$ replica $x$ relays it to the rest of $R$. Since $A\ge 4\Delta$, every replica in $R$, and every replica in $\mathcal{C}_M\cap H$ by the direct argument, includes a proposal referencing $M$ in its own close proposal: every replica in $N_M$ is a carrier. With $|N_M|\ge f+1$, the counting above gives $\Ptr(W_e)\succeq M$ for every certified block.

\textbf{Impossibility.} First, epoch $e-1$ is driven to produce two certified blocks; this is possible for every $R$, including $R=H$. A Byzantine replica withholds its own close proposal for epoch $e-1$ until every honest replica has cast its first vote for the common witness built from the other $3f$ close proposals; these $2f+1$ honest votes certify $B_{lo}$. The withheld close proposal is then delivered, every honest replica extends its witness by an elementary step and votes again, and the resulting certificate defines $B_{hi} \succ B_{lo}$. Set $M = B_{hi}$. Before epoch $e$ opens anywhere, the adversary delivers a message carrying $VC(B_{hi})$ to exactly the clients of the family, so that each references $M = B_{hi}$ in its proposal; every other compliant client first delivers $VC(B_{lo})$, through the openings below.

For epoch $e$, let $K$ consist of the Byzantine replicas and the honest replicas outside $N_M$; since $|N_M|\le f$, the set $K$ contains at least $f+1$ honest replicas, hence at least $2f+1$ replicas in total. Every link from a family client or from $N_M$ toward an honest member of $K$ is adversarial: an honest member of $K$ lies in no reliable set of the family and, whenever $\mathcal{C}_M\cap R\neq\emptyset$, it lies outside $R\subseteq N_M$; if instead $\mathcal{C}_M\cap R=\emptyset$, then $N_M=\mathcal{C}_M\cap H$ is disjoint from $R$, and the delayed links again lie outside $R$. A Byzantine client opens the epoch for $K$ only, with a valid empty-payload proposal referencing $B_{lo}$; all messages from the family clients and from $N_M$ toward $K$ are delayed until the voting of $K$ has ended. The family clients' messages to their reliable sets are delivered within $\Delta$, but they reach only replicas in $\mathcal{C}_M\cap H\subseteq N_M$, whose messages toward $K$ are delayed. The honest members of $K$ exchange close proposals promptly --- the adversary can make adversarial links arbitrarily fast --- each observes the common set of close proposals of the replicas in $K$, none of which contains a proposal referencing a block above $B_{lo}$, waits until time $2A$ after opening, and votes for the common witness $W_e$ with $\Ptr(W_e)=B_{lo}\prec M$. Completed by Byzantine votes, these at least $2f+1$ votes form a certificate. The Byzantine client then opens epoch $e+1$ for $K$, ending its voting phase for epoch $e$; all delays are finite.

\subsection{Proof of Lemma~\ref{lem:partial-inclusion}}

\textbf{Part (i).} Label the honest replicas $r,h_1,\dots,h_{2f}$ and the Byzantine replicas $b_1,\dots,b_f$. In each construction below we describe an event on nature's draws; the event has positive probability by full support, non-degeneracy, and independence.

Suppose first that $A<2\bar{p}-\underline{p}$ and the epoch is opened by a client broadcasting a valid proposal to every replica at a common time, normalized to $0$. Consider the event that the delay of the opening proposal toward $r$ falls within $[\bar{p}-\varepsilon,\bar{p}]$ while its delays toward all other replicas fall within $[\underline{p},\underline{p}+\varepsilon]$: every replica except $r$ opens at about $\underline{p}$ and closes at about $\underline{p}+A$, while $r$ opens at about $\bar{p}$ and closes at about $\bar{p}+A$. The compliant client first delivers a vote certificate of epoch $e-1$ through the fastest re-broadcast of the opening proposal, at about $\underline{p}+\rho$, where $\rho$ is the delay of that re-broadcast, and broadcasts $P$ then. Suppose the delay of $P$ toward $r$ falls near $\underline{p}$ and its delays toward all other replicas fall near $\bar{p}$. Then $P$ reaches $r$ before the close boundary of $r$ provided $\underline{p}+\rho+\underline{p}\le\bar{p}+A$, and reaches every other replica after its own close boundary provided $\underline{p}+\rho+\bar{p}>\underline{p}+A$. Both hold for any $\rho$ with
\[
\max\{\underline{p},\,A-\bar{p}\} \;<\; \rho \;\le\; \min\{\bar{p},\,A+\bar{p}-2\underline{p}\},
\]
an interval that is nonempty since $3\underline{p}-\bar{p}<A<2\bar{p}$. On the corresponding event, $\CP_r$ contains $P$ and no other close proposal does; if $P$ reaches $r$ before the opening proposal does, then $P$ itself opens the epoch at $r$. At its close boundary, each Byzantine replica sends its close proposal, which omits $P$. On the event that the close proposals of $h_1,\dots,h_{2f},b_1,\dots,b_f$ are delivered among the replicas other than $r$ with delays near $\underline{p}$, while $\CP_r$ --- sent at about $\bar{p}+A$ --- reaches every replica with delay near $\bar{p}$, hence at about $2\bar{p}+A$, each of $h_1,\dots,h_{2f}$ observes all $3f$ close proposals other than $\CP_r$, and not $\CP_r$, by the time it can vote: the arrival of $\CP_r$ exceeds the gate at $\underline{p}+2A$ precisely because $A<2\bar{p}-\underline{p}$. Whether at the gate or at the arrival of the last of the $3f$ close proposals, each therefore votes for the common witness $W^{*}$ built from the $3f$ close proposals other than $\CP_r$; the Byzantine replicas vote for $W^{*}$ as well. These at least $2f+1$ votes form $VC(W^{*})$, and, since $P\notin\mathcal{P}(W^{*})$, $\mathsf{tx}(P)\notin B(W^{*})$.

Suppose instead that a Byzantine client, holding a vote certificate of epoch $e-1$, opens the epoch by sending a valid empty-payload proposal to every replica except $r$; we show the same conclusion for every $A<2\bar{p}$. Consider the event that the opening proposal reaches every recipient with delay in $[\underline{p},\underline{p}+\varepsilon]$, while every re-broadcast of it toward $r$ has delay in $[\bar{p}-\varepsilon,\bar{p}]$: every replica except $r$ opens at about $\underline{p}$, while $r$ opens at about $\underline{p}+\bar{p}$ --- or earlier, if the compliant client's proposal reaches it first. Withholding the opening from $r$ disperses the openings by up to $\bar{p}$ rather than $\bar{p}-\underline{p}$, and this wider dispersion is what extends the range of $A$. The compliant client broadcasts $P$ at about $\underline{p}+\rho$, where $\rho\in[\underline{p},\bar{p}]$ is the delay of the fastest re-broadcast toward it. Suppose the delays of $P$ toward $h_1,\dots,h_{2f}$ fall near $\bar{p}$ and $\rho>A-\bar{p}$: then $P$ reaches each of them after its close boundary, and no close proposal other than $\CP_r$ contains $P$. Suppose further that the delay $\delta$ of $P$ toward $r$ satisfies $\rho+\delta\le\bar{p}+A$: then $P$ reaches $r$ before the close boundary of $r$, at about $\underline{p}+\bar{p}+A$, or opens $r$ itself if it arrives before the re-broadcasts; in either case $P\in\CP_r$. When $A\ge\underline{p}$, take $\delta$ near $\underline{p}$; when $A<\underline{p}$, the hypothesis $A>3\underline{p}-\bar{p}$ forces $\bar{p}>2\underline{p}$, and $\rho,\delta$ near $\underline{p}$ give $\rho+\delta\le\bar{p}$. Together with $\rho>A-\bar{p}$, which is compatible with $\rho\le\bar{p}$ precisely when $A<2\bar{p}$, all constraints are satisfied on an event of positive probability. At its close boundary, each Byzantine replica sends its close proposal, which omits $P$; on the event that all close proposals other than $\CP_r$ are delivered among the replicas other than $r$ with delays near $\underline{p}$, while $\CP_r$ --- sent at about $\underline{p}+\bar{p}+A$ --- reaches every replica with delay near $\bar{p}$, hence at about $\underline{p}+2\bar{p}+A$, the arrival of $\CP_r$ exceeds the gate at $\underline{p}+2A$, since $A<2\bar{p}$, and $VC(W^{*})$ forms exactly as above.

In either construction, all remaining deliveries complete in finite time, and the adversary delays the opening of epoch $e+1$ until they do. Upon observing $\CP_r$, every honest replica extends its view by an elementary step and votes for the common witness $W^{**}$ built from all $3f+1$ close proposals, with $W^{*}\prec W^{**}$; the $2f+1$ honest votes form $VC(W^{**})$. Since $P\in\mathcal{P}(W^{**})$, the client does not equivocate, client signatures are unforgeable, and monotone validity keeps $\mathsf{tx}(P)$ valid relative to $\Ptr(W^{**})$, we conclude $\mathsf{tx}(P)\in B(W^{**})$ and $B(W^{*})\prec B(W^{**})$.

\textbf{Part (ii).} If the Byzantine replicas send no close proposals, the only close proposals in existence are the $2f+1$ honest ones, and every votable witness must contain close proposals from at least $2f+1$ distinct senders: every witness that receives an honest vote contains every honest close proposal. If some honest close proposal contains $P$, then $P\in\mathcal{P}$ of every such witness, since honest senders are never flagged and $P$ is valid by compliance; the client does not equivocate and monotone validity keeps $\mathsf{tx}(P)$ valid relative to every pointer, so $\mathsf{tx}(P)$ belongs to every certified block. If no honest close proposal contains $P$, then $P$ belongs to $\mathcal{P}$ of no witness, and $\mathsf{tx}(P)$ belongs to no certified block.

\section{Simulation details}\label{app:simulation}
Suppose the transaction of a compliant client is contained in the certified block $B(W_{\hat e})$ of epoch $\hat e$. In each later epoch the client broadcasts an empty-payload proposal relaying the highest vote certificate it has delivered whose block extends $B(W_{\hat e})$, upon first delivering any vote certificate of the previous epoch. If this proposal reaches every honest replica before its close boundary, then every certified witness of the epoch contains it, hence every certified block of the epoch extends a block that carries the transaction; by Lemmas~\ref{lem:same-epoch-monotonicity} and~\ref{lem:cross-epoch-monotonicity} the transaction then belongs to every active fork, and it does so permanently, whatever the later draws.

Whether the relay succeeds in a given epoch is determined by nature's draws, and we now define the resulting probability. Fix the behavior of the Byzantine processes and one realization of the delays. The client's proposal then reaches each replica at a fixed time, while raising $A$ postpones every close boundary. A larger $A$ can therefore only add replicas whose close proposal carries the proposal, never remove them, so no witness loses it. Raising $A$ also postpones the start of voting, so first witnesses can only grow richer; both effects push toward inclusion. Hence in each realization there is a critical value $\tau$ such that the transaction is contained in every certified block of the epoch if and only if $A>\tau$. We define
\[
q(A)\;=\;\Pr\bigl[\,\tau<A\,\bigr],
\]
the probability that the transaction is included in \emph{all} certified blocks of the epoch. Written this way, $q$ is a distribution function, and three of its properties determine how $A$ should be chosen. Note that $q(A)$ is weakly increasing in $A$: the events $\{\tau<A\}$ are nested. Also, $q(A)=1$ for $A>2\bar{p}$, because $\tau\le 2\bar{p}$ in every realization. Finally, $q(A)>0$ for every $A>3\underline{p}-\bar{p}$: with positive probability one replica opens the epoch early and informs the client quickly, while at least $f+1$ honest replicas open late and hence close late enough to receive the client's proposal in time, so that Lemma~\ref{prop:inclusion-safety} applies.

The choice of $A$ is therefore a trade-off. Since $A$ governs the duration of an epoch, a small $A$ means fast settlement, and this is the reason not to set $A$ at the conservative value $2\bar{p}$ at which inclusion in all certified blocks is certain. But a smaller $A$ means a smaller $q(A)$, and hence a higher chance that the transaction is included in only some of the epoch's certified blocks --- the partial-inclusion event of Lemma~\ref{lem:partial-inclusion}. The second path is what makes a smaller $A$ tolerable: it does not restore immediacy, but it lets a client whose transaction was only partially included converge to inclusion in all active forks, at the cost of waiting for an epoch in which the draws are favorable. How steeply $q$ rises with $A$, and how quickly the relay converges, are quantitative questions, and we turn to simulation to answer them.

\begin{figure}[t]
\centering
\resizebox{.8\textwidth}{!}{
\begin{tikzpicture}[x=1.55cm, y=4.0cm, >={Latex[length=2mm]}]
\draw[->, thick] (0,0) -- (7.05,0) node[below left=2pt and -8pt, font=\small] {$A$};
\draw[->, thick] (0,0) -- (0,1.12);
\foreach \yy in {0,0.25,0.5,0.75,1} {
  \draw (-0.04,\yy) -- (0.04,\yy);
  \node[left, font=\scriptsize] at (-0.05,\yy) {\yy};}
\foreach \xx in {1,2,3,4,5,6} {
  \draw (\xx,-0.012) -- (\xx,0.012);
  \node[below, font=\scriptsize] at (\xx,-0.02) {\xx};}
\draw[dashed, black!45] (1,0) -- (1,1.06) node[above, font=\scriptsize] {$\underline{p}$};
\draw[dashed, black!45] (6,0) -- (6,1.06) node[above, font=\scriptsize] {$2\bar{p}$};
\draw[green!55!black, very thick] plot coordinates {(0.25,0.000) (0.50,0.001) (0.75,0.005) (1.00,0.017) (1.25,0.046) (1.50,0.096) (1.75,0.174) (2.00,0.296) (2.25,0.459) (2.50,0.635) (2.75,0.785) (3.00,0.902) (3.25,0.959) (3.50,0.985) (3.75,0.996) (4.00,0.999) (4.25,1.000) (4.50,1.000) (4.75,1.000) (5.00,1.000) (5.25,1.000) (5.50,1.000) (5.75,1.000) (6.00,1.000) (6.25,1.000) (6.50,1.000)};
\draw[blue!65!black, very thick] plot coordinates {(0.25,0.000) (0.50,0.000) (0.75,0.002) (1.00,0.011) (1.25,0.030) (1.50,0.067) (1.75,0.117) (2.00,0.138) (2.25,0.134) (2.50,0.101) (2.75,0.063) (3.00,0.026) (3.25,0.007) (3.50,0.001) (3.75,0.001) (4.00,0.000) (4.25,0.000) (4.50,0.000) (4.75,0.000) (5.00,0.000) (5.25,0.000) (5.50,0.000) (5.75,0.000) (6.00,0.000) (6.25,0.000) (6.50,0.000)};
\draw[red!70!black, very thick] plot coordinates {(0.25,1.000) (0.50,0.999) (0.75,0.993) (1.00,0.972) (1.25,0.923) (1.50,0.837) (1.75,0.709) (2.00,0.567) (2.25,0.407) (2.50,0.264) (2.75,0.152) (3.00,0.072) (3.25,0.034) (3.50,0.014) (3.75,0.004) (4.00,0.001) (4.25,0.000) (4.50,0.000) (4.75,0.000) (5.00,0.000) (5.25,0.000) (5.50,0.000) (5.75,0.000) (6.00,0.000) (6.25,0.000) (6.50,0.000)};
\draw[black, thick, densely dashed] plot coordinates {(0.25,0.000) (0.50,0.001) (0.75,0.005) (1.00,0.018) (1.25,0.048) (1.50,0.103) (1.75,0.197) (2.00,0.343) (2.25,0.530) (2.50,0.707) (2.75,0.838) (3.00,0.926) (3.25,0.966) (3.50,0.986) (3.75,0.996) (4.00,0.999) (4.25,1.000) (4.50,1.000) (4.75,1.000) (5.00,1.000) (5.25,1.000) (5.50,1.000) (5.75,1.000) (6.00,1.000) (6.25,1.000) (6.50,1.000)};
\draw[green!55!black, very thick] (3.9,0.50) -- (4.25,0.50) node[right, black, font=\scriptsize] {recovery this epoch, $q(A)$};
\draw[black, thick, densely dashed] (3.9,0.40) -- (4.25,0.40) node[right, font=\scriptsize] {inclusion safety achieved};
\draw[red!70!black, very thick] (3.9,0.30) -- (4.25,0.30) node[right, black, font=\scriptsize] {fork abandoned};
\draw[blue!65!black, very thick] (3.9,0.20) -- (4.25,0.20) node[right, black, font=\scriptsize] {active, not recovered};
\end{tikzpicture}
}
\caption{\small Simulation of the relay epoch in the stochastic model ($f=2$, delays i.i.d.\ uniform on $[\underline{p},\bar{p}]=[1,3]$, $6{,}000$ epochs per point). For each close boundary $A$, the curves report the probability that every certified block of the epoch carries the relayed transaction, which is $q(A)$; that some but not all do; and that none does, so that the fork carrying the transaction is abandoned. The dashed curve is the probability that recovery arrives before the fork is abandoned when epochs repeat independently, which equals the probability with which the protocol achieves inclusion safety for the transaction (Definition~\ref{def:inclusion-safety}). Here $3\underline{p}-\bar{p}=0$, so $q(A)$ is positive for every positive $A$.}
\label{fig:recovery-sim}
\end{figure}
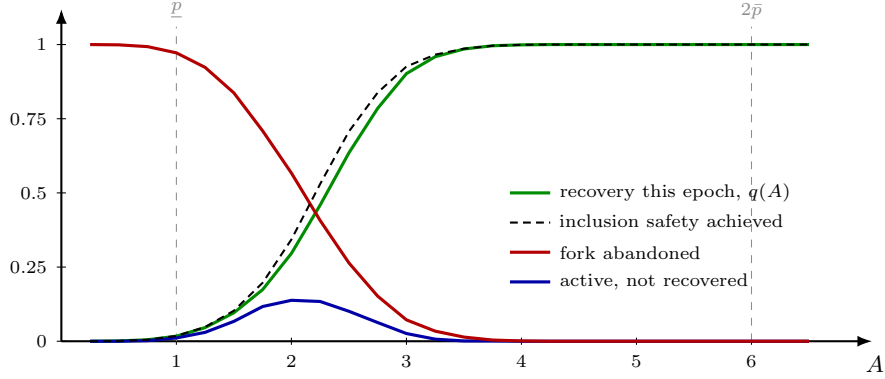

Figure~\ref{fig:recovery-sim} quantifies these dynamics. We simulate the relay epoch with $f=2$ and delays drawn uniformly from $[\underline{p},\bar{p}]=[1,3]$, under the representative adversary described below. For each value of $A$, the figure reports the probability that the epoch ends in recovery, which is $q(A)$ (every certified block carries the transaction), in partial survival (some do), or with the fork abandoned (none does). Two features stand out. First, recovery is almost certain already at $A\approx 4$, well below $2\bar{p}=6$, because the client learns a vote certificate through the \emph{fastest} of the honest re-broadcasts, whose delay concentrates near $\underline{p}$. Second, the intermediate state --- fork active but not recovered --- is thin at every $A$: epochs essentially resolve one way or the other, and a client that falls back on the relay rarely waits more than one epoch. The dashed curve aggregates across epochs: it is the probability that recovery arrives before the fork is abandoned when epochs repeat independently, which is precisely the probability with which the protocol achieves inclusion safety for the transaction (Definition~\ref{def:inclusion-safety}). Below $A=2\bar{p}$, inclusion safety in the stochastic model is therefore probabilistic, with the probability shown; certainty returns at $A\ge 2\bar{p}$, where the first path is always active.

The simulation implements one relay epoch of the stochastic model of Section~\ref{sec:alternating}, with $n=3f+1=7$ replicas, $f=2$ Byzantine replicas, and message delays drawn i.i.d.\ uniformly from $[\underline{p},\bar{p}]=[1,3]$. The close boundary $A$ ranges over a grid from $0.25$ to $6.5$, with $6{,}000$ simulated epochs per grid point. Throughout, the relayed transaction is valid against every block of the epoch preceding its submission.

The Byzantine processes behave as follows. A Byzantine client opens the epoch: holding a vote certificate of the previous epoch whose block omits the transaction, it sends a valid proposal to every replica at the start of the epoch, so each honest replica opens upon delivery, closes one interval $A$ later, and starts voting one further interval $A$ after that. The compliant client learns a vote certificate of the previous epoch only through the re-broadcast of that opening proposal --- the hidden-certificate scenario --- and broadcasts its relay upon delivery. The $f$ Byzantine replicas omit the relay from their close proposals, send those close proposals at the earliest honest close boundary, and vote for every witness that omits the relay. Voting is truncated at the first vote certificate plus $A$, a conservative approximation of the close boundary of epoch $e+1$.

The simulated adversary is representative rather than worst-case: the Byzantine processes do not equivocate and do not inject proposals referencing higher blocks of the previous epoch. One consequence is worth noting. Under this adversary, the openings are dispersed by at most $\bar{p}-\underline{p}$, so once $A \ge 2\bar{p}-\underline{p}$ every close proposal reaches every replica before its first vote, and partial inclusion is impossible. A worst-case adversary can do better: by withholding the opening proposal from one replica, it disperses the openings by up to $\bar{p}$ and sustains partial inclusion for every $A<2\bar{p}$, as in the proof of Lemma~\ref{lem:partial-inclusion}.

Each simulated epoch ends in one of three states: recovery (every certified block carries the transaction), partial survival (some do), or abandonment (none does). Relating these outcomes to Definition~\ref{def:inclusion-safety}: recovery in some epoch means that the protocol achieves inclusion safety for the transaction --- immediate inclusion held in the epoch in which the transaction was submitted, where it entered some certified block, and the recovery epoch supplies the $e^\star$ of the eventual-agreement clause. Abandonment is instead a failure of that clause: the surviving active forks contain no block that includes the transaction and, by the invalid-not-included clause, never will. The dashed curve of Figure~\ref{fig:recovery-sim} is therefore the probability with which the protocol achieves inclusion safety when epochs repeat independently.

\end{document}